\documentclass[a4paper,12pt]{article}
\usepackage{epsfig}
\usepackage{color}
\usepackage[numbers,sort&compress]{natbib}

\renewcommand{\baselinestretch}{1.5}

\newlength{\dinwidth}
\newlength{\dinmargin}
\begin{document}
\title{ Studying Semi-leptonic $b\to (s,d) \nu \bar{\nu}$ Decays in the MSSM without R-parity }
  \author{C. S. Kim\thanks{cskim@yonsei.ac.kr} ~and
  Ru-Min Wang\thanks{ruminwang@cskim.yonsei.ac.kr}
  \\
  {\small  \it Department of Physics and IPAP, Yonsei University, Seoul 120-479, Korea}
  }

 \maketitle \vspace{1.5cm}

\begin{abstract}

\noindent We present a complete study of R-parity violating
supersymmetric effects in  thirteen exclusive and inclusive
semi-leptonic $b \to (s,d) \nu \bar{\nu}$ decays, including
$B^+_u\to K^{(*)+}\nu\bar{\nu}$, $B^0_d\to K^{(*)0}\nu\bar{\nu}$,
$B^0_s\to \phi\nu\bar{\nu}$, $B^0_d\to \pi^0(\rho^0)\nu\bar{\nu}$,
$B^+_u\to \pi^+(\rho^+)\nu\bar{\nu}$, $B^0_s\to
K^{(*)0}\nu\bar{\nu}$ and $B \to X_{s,d} \nu \bar{\nu}$ decay modes,
and we find those thirteen  modes are very sensitive to the
constrained R-parity violating couplings. We derive stringent bounds
on relevant R-parity violating couplings, which are based on all
existent experimental upper limits of involved semi-leptonic decays.
In addition, we also investigate the sensitivities of the branching
ratios and di-neutrino invariant mass spectra to the survived
R-parity violating coupling spaces. Since the experimental bounds
would become much better soon through Super-B, we expect that future
experiments will greatly strengthen our bounds.

\vspace{1.5cm} \noindent {PACS Numbers: 11.30.Fs, 13.20.He,
12.60.Jv}

\end{abstract}

\newpage
\section{Introduction}

The flavor changing neutral current (FCNC) processes are forbidden
at tree level and occur at the lowest order only through one-loop
diagrams in the standard model (SM).  On the other hand, FCNC processes are very sensitive to possible
new physics (NP) scenarios beyond the SM, and provide a unique source of
constraints on some NP scenarios which predict a large change of
these processes. And thus, the measurement of these processes has a very
good chance to reveal NP beyond the SM. Therefore, they are
widely recognized as a powerful tool to make stringent test of the SM.

Rare $B$ decays with a $\nu\bar{\nu}$ pair in the final state, as
such FCNC examples, can be investigated through the large missing
energy associated with the two neutrinos. On the other hand,
experimental search of semi-leptonic $b\to (s,d)\nu \bar{\nu}$
decays is a hard task.  At present, only the upper bounds have been
set by the B{\footnotesize A}B{\footnotesize AR}, Belle, DELPHI and
ALEPH collaborations. We summarize here experimental upper limits
for semi-leptonic $b\to s\nu \bar{\nu}$ and $b\to d\nu \bar{\nu}$
decays at the $90\%$ C.L. in Eq. (\ref{eq:bsvv}) and Eq.
(\ref{eq:bdvv}), respectively.
\begin{eqnarray}
&&\mathcal{B}(B^0_d\to K^0\nu\bar{\nu})<160\times10^{-6}
~\mbox{\cite{:2007zk}}, ~~~~~\mathcal{B}(B^+_u\to
K^+\nu\bar{\nu})<14\times10^{-6}~\mbox{\cite{:2007zk}},\nonumber\\
&& \mathcal{B}(B^0_d\to K^{*0}\nu\bar{\nu})<120\times10^{-6}
~\mbox{\cite{:2008fr}},~~~~
 \mathcal{B}(B^+_u\to
K^{*+}\nu\bar{\nu})<80\times10^{-6}~\mbox{\cite{:2008fr}},\nonumber\\
 &&\mathcal{B}(B^0_s\to
\phi\nu\bar{\nu})<5400\times10^{-6}~\mbox{\cite{Adam:1996ts}},~~~~
 \mathcal{B}(B\to X_s\nu\bar{\nu})<640\times10^{-6}~\mbox{\cite{Barate:2000rc}},\label{eq:bsvv}\\
{\rm and}~~~ &&\mathcal{B}(B^0_d\to
\pi^0\nu\bar{\nu})<220\times10^{-6}~\mbox{\cite{:2007zk}},~~~~~
\mathcal{B}(B^+_u\to \pi^+\nu\bar{\nu})<100\times10^{-6}~\mbox{\cite{Aubert:2004ws}},\nonumber\\
&&\mathcal{B}(B^0_d\to
\rho^{0}\nu\bar{\nu})<440\times10^{-6}~\mbox{\cite{:2007zk}},~~~~~
\mathcal{B}(B^+_u\to
\rho^{+}\nu\bar{\nu})<150\times10^{-6}~\mbox{\cite{:2007zk}}.\label{eq:bdvv}
\end{eqnarray}

 Theoretically, $b\to (s,d) \nu \bar{\nu}$ decays are very clean processes, which are
 sensitive to several possible sources of NP \cite{Grossman:1995gt}.
 The NP effects in $b\to (s,d) \nu \bar{\nu}$ decays have been
investigated by many authors (see e.g., Refs.
\cite{Grossman:1995gt,Melikhov:1998ug,Kim:1999waa,Buchalla:2000sk,Aliev:2001in,
Jeon:2006nq,Aliev:2007gr,Blanke:2008yr,Altmannshofer:2009ma}).
 Supersymmetry  is one of the most widely discussed options
of NP, in both its R-parity conserving
 and R-parity violating (RPV) incarnations
\cite{Aulakh:1982yn,Ross:1984yg}. In recent papers
we have presented detailed study of charged
Higgs effects and  RPV effects in rare exclusive $b\to u \ell
\nu_\ell$ \cite{OurRPVstudy1} and $b\to c\bar{c}s(d)$ decays \cite{OurRPVstudy2}.
 In the
minimal supersymmetric standard model (MSSM) \cite{MSSM1,MSSM2} with
R-parity conservation, the new contributions to the $b\to s \nu \bar{\nu}$
transition have been discussed  (for instance, see Refs.
\cite{Bertolini:1990if,Goto:1996dh,Yamada:2007me}).
 In this work, we will
concentrate on RPV effects in the exclusive and inclusive
semi-leptonic $b\to (s,d)\nu \bar{\nu}$ decays. {}From the latest
experimental data given in Eqs. (\ref{eq:bsvv}-\ref{eq:bdvv}) and
the theoretical parameters with uncertainties, we will derive the
new conservative upper limits on the relevant RPV coupling products.
Moreover, we will also investigate how survived RPV coupling spaces
can affect on the branching ratios and di-neutrino invariant mass
($i.e.$ missing mass) spectra in these  semi-leptonic $b\to (s,d)\nu
\bar{\nu}$ decays. We find these observables are still very
sensitive to survived RPV coupling spaces.

Our letter is organized as follows: In Sec. 2, we review the
effective Hamiltonian for $b\to (s,d)\nu \bar{\nu}$ transitions and
define the observables that can in principle be measured in these
decays. In Sec. 3, we deal with the numerical results. We display
the constrained parameter spaces which satisfy all the available
experimental upper limits of the  $b\to (s,d)\nu \bar{\nu}$, and
then, we investigate the sensitivities of the branching ratios and
di-neutrino invariant mass spectra to the survived RPV coupling
spaces in those decays.  We conclude in Sec. 4.

\section{Theoretical Framework}

The  $b\to d_j \nu_{i'} \bar{\nu}_i$ ($j=1,2$ and $i,i'=e,
\mu,\tau$) transitions can be described by the effective Hamiltonian,
\begin{eqnarray}
\mathcal{H}_{\rm eff}(b\to d_j \nu_{i'} \bar{\nu}_i)=C^\nu_L~
\bar{b}\gamma_\mu(1-\gamma_5)d_j
\bar{\nu}_i\gamma^\mu(1-\gamma_5)\nu_{i'}+C^\nu_R~
\bar{b}\gamma_\mu(1+\gamma_5)d_j
\bar{\nu}_{i}\gamma^\mu(1-\gamma_5)\nu_{i'}~. \label{Heff}
\end{eqnarray}
In the SM, $b\to d_j \nu_{i'} \bar{\nu}_i$ proceeds via $W$ box and
$Z$ penguin diagrams, therefore only purely left-handed currents
$\bar{b}\gamma_\mu(1-\gamma_5)d_j
\bar{\nu}_i\gamma^\mu(1-\gamma_5)\nu_{i'}$ are present. The
corresponding left-handed coefficient reads $C^{\nu}_{L,{\rm
SM}}=\frac{G_F\alpha_e}{2\pi\sqrt{2}}V_{td_j}V_{tb}^*
X(x_t)/\mbox{sin}^2\theta_W$ \cite{Inami:1980fz}, where $G_F$ is the
Fermi constant, $\alpha_e$ is the fine structure constant,
$\theta_W$ is the Weinberg angle, and $V_{ij}$ are the CKM matrix
elements. Function  $X(x_t)$ is dominated by the short-distance
dynamics associated with top quark exchange \cite{Buchalla:2000sk},
and has the theoretical uncertainty due to the error of top quark
mass, whose explicit form  can be found in Refs.
\cite{Misiak:1999yg,Buchalla:1998ba}.


In supersymmetric models without R-parity
\cite{Aulakh:1982yn,Ross:1984yg}, extra trilinear RPV terms\footnote{$\hat{L}$ and
$\hat{Q}$ are the SU(2) doublet lepton and quark superfields,
respectively, $\hat{D}^c$ are the singlet superfields, while $i$,
$j$ and $k$ are generation indices and the superscript $c$ denotes a
charge conjugate field.}
$\lambda'_{ijk}\hat{L}_i\hat{Q}_j\hat{D}^c_k$ are allowed in the superpotential
\cite{RPVSW}.  Both left-handed and right-handed currents are
present in $b\to d_j \nu_{i'} \bar{\nu}_i$ transition at the tree
level in these models. Then the corresponding coefficients in Eq.
(\ref{Heff}) are written as
\begin{eqnarray}
C^\nu_L=C^\nu_{L,{\rm SM}}-\sum_k\frac{\lambda'^{*}_{i3k}\lambda'_{i'jk}}{8m^2_{\tilde{d}_{kR}}}~,~~~~~
C^\nu_R=\sum_k\frac{\lambda'^{*}_{ikj}\lambda'_{i'k3}}{8m^2_{\tilde{d}_{kL}}}~.\label{Coff}
\end{eqnarray}
RPV couplings $\lambda'^{*}_{i3k}\lambda'_{i'jk}$  arise  from
right-handed squark exchanges, and
$\lambda'^{*}_{ikj}\lambda'_{i'k3}$ come from left-handed squark
exchanges. Note that the RPV coupling coefficient $\lambda'_{ijk}$
can be a complex in our convention, which is different from Ref.
\cite{Grossman:1995gt}.

From the theoretical point of view, the inclusive semi-leptonic
$b\to q \nu\bar{\nu}$ $(q=s,d)$ decays  are very clean proceses,
since both the perturbative $\alpha_s$ and the non-perturbative
$1/m_b^2$ corrections are known to be small. Their dineutrino
invariant mass distributions are given as following
\begin{eqnarray}
\frac{d\mathcal{B}(B\to X_q
\nu_{i'}\bar{\nu}_i)}{ds_b}&=&\frac{\tau_B
\kappa(0)}{16\pi^3m_b^3}\left(|C_L^\nu|^2+|C_R^\nu|^2\right)\sqrt{\lambda(m_b^2,m_q^2,s_b)}\nonumber\\
&&\times\left[3s_b\left(m_b^2+m_q^2-s_b-4m_bm_q\frac{\rm{Re}(C_L^\nu
C_R^{\nu*})}{|C_L^\nu|^2+|C_R^\nu|^2}\right)+\lambda(m_b^2,m_q^2,s_b)\right],\label{INCdB}
\end{eqnarray}
where $s_b=(p_b-p_q)^2$, $\lambda(a,b,c)=a^2+b^2+c^2-2ab-2ac-2bc$,
and $\kappa(0)=0.83$ represents the QCD correction to the $b\to q
\nu\bar{\nu}$ matrix element
\cite{Grossman:1995gt,Buchalla:1995vs,Bobeth:2001jm}. We have summed
over the neutrino flavors in Eq.(\ref{INCdB}).

In order to compute branching ratios of the exclusive semi-leptonic
$b\to (s,d)\nu\bar{\nu}$ decays, we need the matrix elements of the
effective hamiltonian between the states of the initial $B$ particle
and the final particles $M,\nu,\bar{\nu}$.
 The hadronic matrix elements for $B \to
P$ transition ($P$ is a pseudoscalar meson, $\pi$ or $K$)  can be
parameterized in terms of the form factors $f^P_+(s_B)$ and
$f^P_0(s_B)$ as
\begin{eqnarray}
&&c_P\langle P(p)|\bar{u}\gamma_{\mu}b|B (p_{_{B}})\rangle
=f^P_+(s_B)(p+p_{_{B}})_\mu+\left[f^P_0(s_B)-f^P_+(s_B)\right]\frac{m^2_B-m^2_P}{s_B}q_\mu,\label{BtoPFF}
\end{eqnarray}
where the factor $c_P$ accounts for the flavor content of particles
($c_P=\sqrt{2}$ for $\pi^0$, and $c_P=1$ for $\pi^-,K^-$) and
$s_B=q^2~~(q=p_{_{B}}-p=p_{\nu}+p_{\bar \nu})$. For $B \to V$
transition ($V$ is a vector $K^*$, $\rho$ or $\phi$ meson) can be
written in terms of five form factors
\begin{eqnarray}
c_V\langle
V(p,\varepsilon^{\ast})|\bar{u}\gamma_{\mu}(1-\gamma_5)b|B
(p_{_{B}})\rangle &&=\frac{2V(s_B)}{m_B+m_V}
\epsilon_{\mu\nu\alpha\beta}\varepsilon^{\ast\nu}p_{_B}^{\alpha}p^{\beta}\nonumber\\
&&-i\left[\varepsilon_{\mu}^\ast(m_B+m_V)A_1(s_B)
-(p_{_B}+p)_{\mu}({\varepsilon^\ast}\cdot{p_{_B}})\frac{A_2(s_B)}
{m_B+m_V}\right]\nonumber \\
&&+iq_{\mu}({\varepsilon^\ast}\cdot{p_{_B}})\frac{2m_V}{s_B}
[A_3(s_B)-A_0(s_B)],\label{BtoVFF}
\end{eqnarray}
where $c_V=\sqrt{2}$ for $\rho^0$, $c_V=1$ for $\rho^-,K^{*-},\phi$,
and with the relation
$A_3(s_B)=\frac{m_B+m_V}{2m_V}A_1(s_B)-\frac{m_B-m_V}{2m_V}A_2(s_B)$.

In terms of the effective Hamiltonian shown in Eq. (\ref{Heff}) and
the relevant form factors given in Eqs. (\ref{BtoPFF}-\ref{BtoVFF}),
the di-neutrino invariant mass distributions for  $B \to
P\nu\bar{\nu}$ and $B \to V\nu\bar{\nu}$ decays can be written as
\cite{Melikhov:1998ug,Colangelo:1996ay}
\begin{eqnarray}
\frac{d\mathcal{B}(B \to P\nu_{i'}\bar{\nu}_i)}{ds_B}
&=&\left|C^\nu_L+C^\nu_R\right|^2
\frac{\tau_Bm_B^3}{2^5\pi^3c_P^2}\lambda_P^{3/2}(s_B)\left[f^{P}_+(s_B)\right]^2,\\
 \frac{d\mathcal{B}(B \to
V\nu_{i'}\bar{\nu}_i)}{ds_B} &=&\left|C^\nu_L+C^\nu_R\right|^2
\frac{\tau_Bm_B^3}{2^7\pi^3c_V^2}\lambda_V^{1/2}(s_B)
\frac{8s_B\lambda_V(s_B)V^2(s_B)}{(1+\sqrt{r_V})^2}\nonumber\\
&+&\left|C^\nu_L-C^\nu_R\right|^2
\frac{\tau_Bm_B^3}{2^7\pi^3c_V^2}\lambda_V^{1/2}(s_B)\frac{1}{r_V}
\left[(1+\sqrt{r_V})^2(\lambda_V(s_B)+12r_Vs_B)A_1^2(s_B)\frac{}{}\right.\nonumber\\
&+&\left.\frac{\lambda_V^2(s_B)A_2^2(s_B)}{(1+\sqrt{r_V})^2}
-2\lambda_V(s_B)(1-r_V-s_B)A_1(s_B)A_2(s_B)\right],
\end{eqnarray}
where $\lambda_M(s_B)=\lambda(1,r_M,s_B/m_B^2)$ with
$r_M=m^2_M/m_B^2$, and we have summed over the neutrino flavors.

For our numerical results, we use the relevant $B\to P(V)$ form
factors given in \cite{BallZwicky}. However, $B_s\to K$ form factors
are not given in LCSR results \cite{BallZwicky}. After discussions
with authors of Ref. \cite{BallZwicky}, we  obtain them as $
F^{B_s\to K}(s_B)=F^{B_{u,d}\to K}(s_B)\left(\frac{F^{B_s\to
K^*}(s_B)}{F^{B_{u,d}\to K^*}(s_B)}\right).$
 The uncertainties of form
factors at $s_B=0$ induced by $F(0)$  are considered, to be
conservative, we adopt these uncertainties for full $s_B$ range. The
CKM matrix elements are taken from \cite{CKMfit}, and masses and
lifetimes are from Ref. \cite{PDG}.

\section{Numerical Results and Discussions}

In this section, we summarize  our numerical results and analysis in
the semi-leptonic $b\to (s,d)\nu \bar{\nu}$ decays. To be
conservative, we use all input parameters which are varied randomly
within 1$\sigma$ ranges in our numerical results.  We use the
average $\tau_B=(\tau_{B^+}+\tau_{B^0})/2$ for the inclusive decays.
When we study the RPV effects, we consider only one RPV coupling
product contributions at one time, neglecting the interferences
between different RPV coupling products, but keeping their
interferences with the SM amplitude. We assume the masses of
sfermions are 500 GeV. For other values of the sfermion masses, the
bounds on the couplings in this paper can be easily obtained by
scaling them by factor
$\tilde{f}^2\equiv(\frac{m_{\tilde{f}}}{500~\rm{GeV}})^2$.

The transitions $b\to (s~\mbox{or}~d) \nu_{i'} \bar{\nu}_i$ involve
the same set of the RPV coupling products for every generation of
neutrinos: For six semi-leptonic $b\to s \nu_{i'} \bar{\nu}_i$
decays, $B^+_u\to K^{(*)+}\nu\bar{\nu}$, $B^0_d\to
K^{(*)0}\nu\bar{\nu}$, $B^0_s\to \phi\nu\bar{\nu}$ and $B\to
X_s\nu\bar{\nu}$, there are two kinds of RPV coupling products,
$\lambda'^*_{i3k}\lambda'_{i'2k}$ and
$\lambda'^*_{i'k2}\lambda'_{ik3}$, which come from left-handed and
right-handed squark exchanges, respectively. For seven semi-leptonic
$b\to d \nu_{i'} \bar{\nu}_i$ decay modes, $B^0_d\to
\pi^0(\rho^0)\nu\bar{\nu}$, $B^+_u\to \pi^+(\rho^+)\nu\bar{\nu}$,
$B^0_s\to K^{(*)0}\nu\bar{\nu}$ and $B\to X_d\nu\bar{\nu}$, RPV
coupling products $\lambda'^*_{i3k}\lambda'_{i'1k}$ and
$\lambda'^*_{i'k1}\lambda'_{ik3}$ arise from left-handed and
right-handed squark exchanges, respectively.
\begin{figure}[t]
\begin{center}
\includegraphics[scale=1]{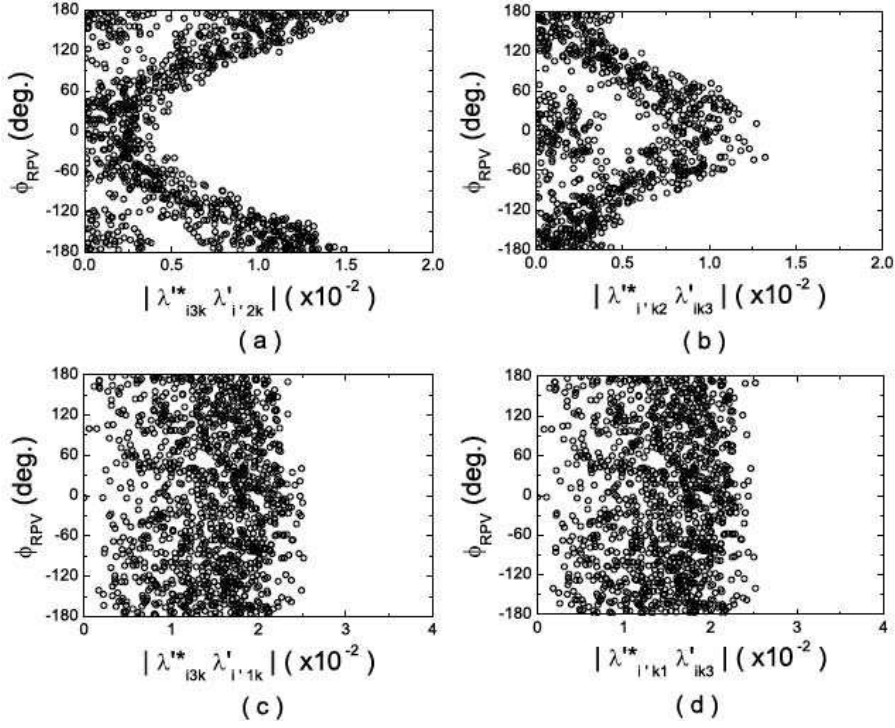}
\end{center}
\vspace{-0.6cm}  \caption{\small Survived parameter spaces  shown
for the relevant RPV coupling products with $500$ GeV sfermion
masses constrained by semi-leptonic (a-b) $b\to s \nu\bar{\nu}$ and
(c-d) $b\to d \nu\bar{\nu}$ decays, respectively, where $\phi_{_{\rm
RPV}}$ denotes the RPV weak phase.} \label{Fig:bounds}
\end{figure}
\begin{table}[t]
\caption{\small Bounds on the relevant RPV coupling products
constrained by semi-leptonic $b\to (s,d) \nu\bar{\nu}$ decays for
500 GeV sfermions. }
\begin{center}
\begin{tabular}{l|l|c}\hline\hline
Couplings &~Our bounds~~~~~~~~~~~[Processes]& Previous
bounds~[Processes]\\\hline
$\left|\lambda'^*_{i3k}\lambda'_{i'2k}\right|$&$\leq1.5\times10^{-2}$~\Bigg[~\parbox{4.45cm}{
\renewcommand{\baselinestretch}{1.1}\scriptsize
$B^+_u\to K^{(*)+}\nu\bar{\nu},
B^0_d\to K^{(*)0}\nu\bar{\nu}\\
B^0_s\to\phi\nu\bar{\nu},B\to X_s
\nu\bar{\nu}$}\Bigg]&$\leq3.5\times10^{-2}~[B \to X_s\nu\bar{\nu}]$
\cite{Grossman:1995gt,Barate:2000rc}\\\hline
$\left|\lambda'^*_{i'k2}\lambda'_{ik3}\right|$&$\leq1.3\times10^{-2}$~\Bigg[~\parbox{4.45cm}{
\renewcommand{\baselinestretch}{1.0}\scriptsize
$B^+_u\to K^{(*)+}\nu\bar{\nu},
B^0_d\to K^{(*)0}\nu\bar{\nu}\\
B^0_s\to\phi\nu\bar{\nu}, B\to X_s
\nu\bar{\nu}$}\Bigg]&$\leq3.5\times10^{-2}~[B\to X_s\nu\bar{\nu}]$
\cite{Grossman:1995gt,Barate:2000rc}\\\hline
$\left|\lambda'^*_{i3k}\lambda'_{i'1k}\right|$&$\leq2.5\times10^{-2}$~\bigg[~\parbox{4.5cm}{
\renewcommand{\baselinestretch}{1.1}\scriptsize
$B^+_u\to \pi^+(\rho^+)\nu\bar{\nu},
B^0_d\to\pi^0(\rho^0)\nu\bar{\nu}$}\bigg]&$\cdots\cdots$\\\hline
$\left|\lambda'^*_{i'k1}\lambda'_{ik3}\right|$&$\leq2.5\times10^{-2}$~\bigg[~\parbox{4.5cm}{
\renewcommand{\baselinestretch}{1.1}\scriptsize
$B^+_u\to \pi^+(\rho^+)\nu\bar{\nu},
B^0_d\to\pi^0(\rho^0)\nu\bar{\nu}$}\bigg]&$\cdots\cdots$\\\hline\hline
\end{tabular}
\end{center}\label{Table:Bounds}
\end{table}
We use the latest experimental upper limits from Refs.
\cite{:2007zk,:2008fr,Adam:1996ts,Barate:2000rc,Aubert:2004ws},
which are listed in Eqs. (\ref{eq:bsvv}-\ref{eq:bdvv}), to constrain
the relevant RPV coupling products. Our bounds on the four RPV
coupling products are demonstrated in Fig. \ref{Fig:bounds}. In Fig.
\ref{Fig:bounds}(a-b), we find that the  $b\to s \nu\bar{\nu}$
decays give quite strong correlation between the moduli and the RPV
weak phases of $\lambda'^*_{i3k}\lambda'_{i'2k}$ and
$\lambda'^*_{i'k2}\lambda'_{ik3}$ coupling products. Fig.
\ref{Fig:bounds}(c-d) show that RPV weak phases of
$\lambda'^*_{i3k}\lambda'_{i'1k}$ and
$\lambda'^*_{i'k1}\lambda'_{ik3}$ are not restricted by current
experimental upper limits of $B^0_d\to \pi^0(\rho^0)\nu\bar{\nu}$
and $B^+_u\to \pi^+(\rho^+)\nu\bar{\nu}$ decays, however,
corresponding moduli are upper limited. The upper limits of the
moduli for the relevant RPV coupling products are summarized in
Table \ref{Table:Bounds}. Our bounds on
$\left|\lambda'^*_{i3k}\lambda'_{i'2k}\right|$ and
$\left|\lambda'^*_{i'k2}\lambda'_{ik3}\right|$, which are mainly
from the semi-leptonic $b\to s \nu \bar{\nu}$ experimental data, are
stronger than ones obtained from the inclusive semi-leptonic $b\to s
\nu \bar{\nu}$ decay \cite{Grossman:1995gt,Barate:2000rc}. We obtain
for the first time the bounds on $\lambda'^*_{i3k}\lambda'_{i'1k}$
and $\lambda'^*_{i'k1}\lambda'_{ik3}$ couplings from the $b\to d \nu
\bar{\nu}$ transitions.

We note that some quadratic RPV coupling combinations,
which contribute to $b\to (s,d) \nu \bar{\nu}$ transitions,  may
also give contributions to $b\to (s,d)\gamma$ and $b \to (s,d)
\ell^+\ell^-~(\ell=e,\mu,\tau)$ processes. The decay $b\to s\gamma$
in the MSSM without R-parity has been shown in
\cite{deCarlos:1996yh} to give weak constraints on relevant RPV
coupling combinations,
$\left|\lambda'^*_{i3k}\lambda'_{i2k}\right|\leq2.25$ and
$\left|\lambda'^*_{ik2}\lambda'_{ik3}\right|\leq 0.87$ with $500$
GeV sfermion masses.  The RPV effects in $b \to (s,d) \ell^+\ell^-$
processes have been studies in Refs.
\cite{Wang:2007sp,Dreiner:2006gu,Xu:2006vk,Saha:2002kt,Jang:1997ry,Chemtob:2004xr,Barbier:2004ez,Barbier:1998fe},
and  some upper limits of their RPV coupling combinations are about
one order of magnitude stronger than ours from $b\to (s,d)
\nu\bar{\nu}$. Here, we list the stronger upper limits from $b \to
(s,d) \ell^+\ell^-$ processes with $500$ GeV sfermions:
$\left|\lambda'^*_{ik2}\lambda'_{ik3}\right|\leq1.2\times 10^{-3}
~(i=1,2)$ \cite{Xu:2006vk},
$\left|\lambda'^*_{ik2}\lambda'_{i'k3}\right|\leq6.7\times 10^{-3}
~(i\neq i')$ \cite{Jang:1997ry},
$\left|\lambda'^*_{1k1}\lambda'_{1k3}\right|\leq2.8\times 10^{-3}$
\cite{Wang:2007sp}, and
$\left|\lambda'^*_{2k1}\lambda'_{2k3}\right|\leq3.3\times 10^{-3}$
\cite{Wang:2007sp}.
%
%
In addition, single bounds of $\lambda'_{ijk}$ are obtained by many
authors (for instance, see Refs.
\cite{Chemtob:2004xr,Barbier:2004ez,Barbier:1998fe,Allanach:2003eb,Allanach:1999bf,Allanach:1999ic,Lebedev,Hirsch:1998mc}).
We also note that some of the single $\lambda'$ couplings can
generate sizable neutrino masses
\cite{Barbier:2004ez,Allanach:2003eb}. Allanach {\em et al}. have
obtained quite strong upper bound $|\lambda'_{ijj}| <10^{-2}$ with
500 GeV sfermions in the RPV mSUGRA model, and  Barbier {\em et al}.
have gotten  $|\lambda'_{i33}| <4.4\times10^{-3}$. Furthermore, the
$\lambda'_{111}$ coupling has been constrained as low as
$|\lambda'_{111}|<1.8\times 10^{-2}$ by neutrino-less double beta
decay \cite{Hirsch:1998mc}. If we now compare our combined bounds
with the products of the single bounds, we find that our combined
bounds are weaker one or two order(s) of magnitude than the products
of the single bounds. However, it also should be noted that the
 parameter spaces of $\lambda'$ from neutrino masses can be evaded
since several other parameters are usually involved in the
extraction of the constraints \cite{Borzumati:2002bf}. Furthermore,
the constraints on $\lambda'$ from neutrino masses would depend on
the explicit neutrino masses models with trilinear couplings only,
bilinear couplings only, or both \cite{Barbier:2004ez}.

 Next, we will first
explore the RPV MSSM effects  by using our constrained RPV parameter
 spaces, and then discuss the RPV effects  after also considering previous
 stronger bounds in the semi-leptonic
 $b\to (s,d) \nu\bar{\nu}$ decays.
Now using the survived RPV parameter spaces shown in Fig.
\ref{Fig:bounds}, we explore the RPV MSSM effects in the
semi-leptonic
 $b\to (s,d) \nu\bar{\nu}$ decays, which satisfy all
experimental upper limits given in Eqs.
(\ref{eq:bsvv}-\ref{eq:bdvv}). Our RPV MSSM predictions within the
theoretical uncertainties of input parameters are  given in Table
\ref{Table:Result}, together with experimental upper limits and the
SM predictions for a convenient comparison.
\begin{table}[b]
\caption{\small The branching ratios of the semi-leptonic $b\to
(s,d) \nu\bar{\nu}$ decays (in units of $10^{-6}$), and $j=2(1)$ for
the $b\to s (d)$ transition.}
\begin{center}{\small
\begin{tabular}{l|c|c|c|c}\hline\hline
~~Observable&{\small Exp. Data
\cite{:2007zk,:2008fr,Adam:1996ts,Aubert:2004ws}}&{\small SM
Predictions}&{\small MSSM
w/$\lambda'^*_{i3k}\lambda'_{i'jk}$}&{\small MSSM
w/$\lambda'^*_{i'kj}\lambda'_{ik3}$}\\\hline
%
$\mathcal{B}(B^0_d\to
K^0\nu\bar{\nu})$&$<160$&$[3.48,6.55]$&$[0.14,13.14]$&$[0.14,13.07]$\\\hline
$\mathcal{B}(B^+_u\to
K^+\nu\bar{\nu})$&$<14$&$[3.75,7.04]$&$[0.15,14.00]$&$[0.15,14.00]$\\\hline
$\mathcal{B}(B^0_d\to
K^{*0}\nu\bar{\nu})$&$<120$&$[6.98,15.19]$&$[0.21,46.14]$&$[5.16,74.66]$\\\hline
$\mathcal{B}(B^+_u\to
K^{*+}\nu\bar{\nu})$&$<80$&$[7.55,16.35]$&$[0.22,49.33]$&$[5.55,80.00]$\\\hline
$\mathcal{B}(B^0_s\to
\phi\nu\bar{\nu})$&$<5400$&$[8.89,18.85]$&$[0.36,56.48]$&$[5.56,161.17]$\\\hline
$\mathcal{B}(B \to
X_s\nu\bar{\nu})$&$<640$&$[31.15,48.94]$&$[2.09,142.30]$&$[31.65,282.06]$\\\hline\hline
%
$\mathcal{B}(B^0_d\to
\pi^0\nu\bar{\nu})$&$<220$&$[0.05,0.12]$&$[0.07,47.00]$&$[0.01,46.73]$\\\hline
$\mathcal{B}(B^+_u\to
\pi^+\nu\bar{\nu})$&$<100$&$[0.11,0.25]$&$[0.14,100.00]$&$[0.02,100.00]$\\\hline
$\mathcal{B}(B^0_s\to
K^{0}\nu\bar{\nu})$&$\cdots\cdots$&$[0.11,0.43]$&$[0.10,165.05]$&$[0.04,166.20]$\\\hline
$\mathcal{B}(B^0_d\to
\rho^{*0}\nu\bar{\nu})$&$<440$&$[0.10,0.29]$&$[0.11,70.48]$&$[0.12,70.48]$\\\hline
$\mathcal{B}(B^+_u\to
\rho^{*+}\nu\bar{\nu})$&$<150$&$[0.22,0.62]$&$[0.24,150.00]$&$[0.26,150.00]$\\\hline
$\mathcal{B}(B^0_s\to
K^{*0}\nu\bar{\nu})$&$\cdots\cdots$&$[0.24,0.62]$&$[0.30,245.25]$&$[0.19,238.58]$\\\hline
$\mathcal{B}(B\to
X_d\nu\bar{\nu})$&$\cdots\cdots$&$[1.17,2.23]$&$[1.62,907.06]$&$[1.63,932.43]$\\\hline\hline
\end{tabular}}
\end{center}\label{Table:Result}
\end{table}
 In
Table \ref{Table:Result}, the second and third columns give the
experimental upper limits and the SM predictions, respectively, the
forth column lists the effects of left-handed squark exchange
coupling $\lambda'^*_{i3k}\lambda'_{i'jk}$, and  the last column
summaries the effects of coupling $\lambda'^*_{i'kj}\lambda'_{ik3}$
due to right-handed squark exchange. Main theoretical uncertainties
of the SM predictions arise from the CKM matrix elements, Wilson
coefficient and hadronic transition form factors(only for the
exclusive decays). Comparing with experimental upper limits and the
SM predictions, we find some salient features of numerical results
of the RPV effects listed in Table \ref{Table:Result}.
\begin{itemize}
\item[\textcircled{\scriptsize 1}]
RPV coupling $\lambda'^*_{i3k}\lambda'_{i'2k}$ is only constrained
by the experimental upper limit of $\mathcal{B}(B^+_u\to
K^{+}\nu\bar{\nu})$,  and bounds on this coupling constant obtained
from other exclusive $b\to s \nu\bar{\nu}$ decays and inclusive
$B\to X_s\nu\bar{\nu}$ are weaker than one obtained from $B^+_u\to
K^{+}\nu\bar{\nu}$ decay. Comparing with the SM predictions, we find
contributions of $\lambda'^*_{i3k}\lambda'_{i'2k}$ coupling could
enlarge the allowed ranges of all relevant branching ratios, their
upper limits are increased two or three times, and their lower
limits are reduced more than  one order.
\item[\textcircled{\scriptsize 2}]
The restrictions of $\lambda'^*_{i'k2}\lambda'_{ik3}$ come from  the
experimental upper limits of $\mathcal{B}(B^+_u\to
K^{+}\nu\bar{\nu})$ and $\mathcal{B}(B^+_u\to K^{*+}\nu\bar{\nu})$.
The $\lambda'^*_{i'k2}\lambda'_{ik3}$ coupling effects are same as
the effects of $\lambda'^*_{i3k}\lambda'_{i'2k}$ coupling in
$B^0_d\to K^{0}\nu\bar{\nu}$ and $B^+_u\to K^{+}\nu\bar{\nu}$
decays. For $B^0_d\to K^{*0}\nu\bar{\nu}$, $B^+_u\to
K^{*+}\nu\bar{\nu}$, $B^0_s\to\phi\nu\bar{\nu}$ and $B\to X_s
\nu\bar{\nu}$ decays, $\lambda'^*_{i'k2}\lambda'_{ik3}$ coupling
could  obviously increase the allowed upper limits of these
branching ratios.
\item[\textcircled{\scriptsize 3}]
RPV couplings $\lambda'^*_{i3k}\lambda'_{i'1k}$ and
$\lambda'^*_{i'k1}\lambda'_{ik3}$ are constrained by the
experimental upper limits of $\mathcal{B}(B^+_u\to
\pi^{+}\nu\bar{\nu})$ and $\mathcal{B}(B^+_u\to
\rho^{+}\nu\bar{\nu})$.  All allowed upper limits of the relevant
branching ratios could be significantly increased by both
$\lambda'^*_{i3k}\lambda'_{i'1k}$ and
$\lambda'^*_{i'k1}\lambda'_{ik3}$ couplings. The upper bounds of RPV
predictions for $\mathcal{B}(B^0_d\to
\pi^{0}(\rho^{0})\nu\bar{\nu})$ are about 6 times stronger than
existing experimental limits. The allowed lower limits of
$\mathcal{B}(B^0_d\to \pi^{0}\nu\bar{\nu})$, $\mathcal{B}(B^+_u\to
\pi^{+}\nu\bar{\nu})$
 and
$\mathcal{B}(B^0_s\to K^{0}\nu\bar{\nu})$ could be evidently
decreased by $\lambda'^*_{i'k1}\lambda'_{ik3}$ coupling.
\end{itemize}

Next we want to illustrate briefly the sensitivities of relevant
observables to RPV couplings. To this end, for each RPV coupling
product, we can present the correlations of di-neutrino invariant
mass spectra and branching ratios within the constrained parameter
space displayed in Fig. \ref{Fig:bounds} by two-dimensional scatter
plots. The RPV coupling $\lambda'^*_{i3k}\lambda'_{i'jk}$ or
$\lambda'^*_{i'kj}\lambda'_{ik3}$ contributions to these
semi-leptonic $B_d$, $B_u$ and $B_s$ decays are very similar to each
other. So we will take an example for $B\to X_{s,d}\nu\bar{\nu}$,
$K^{+}(K^{*+})\nu\bar{\nu}$, $\pi^{+}(\rho^+)\nu\bar{\nu}$
 decays to illustrate the sensitivities of
quantities to RPV couplings.

\begin{figure}[t]
\begin{center}
\includegraphics[scale=1]{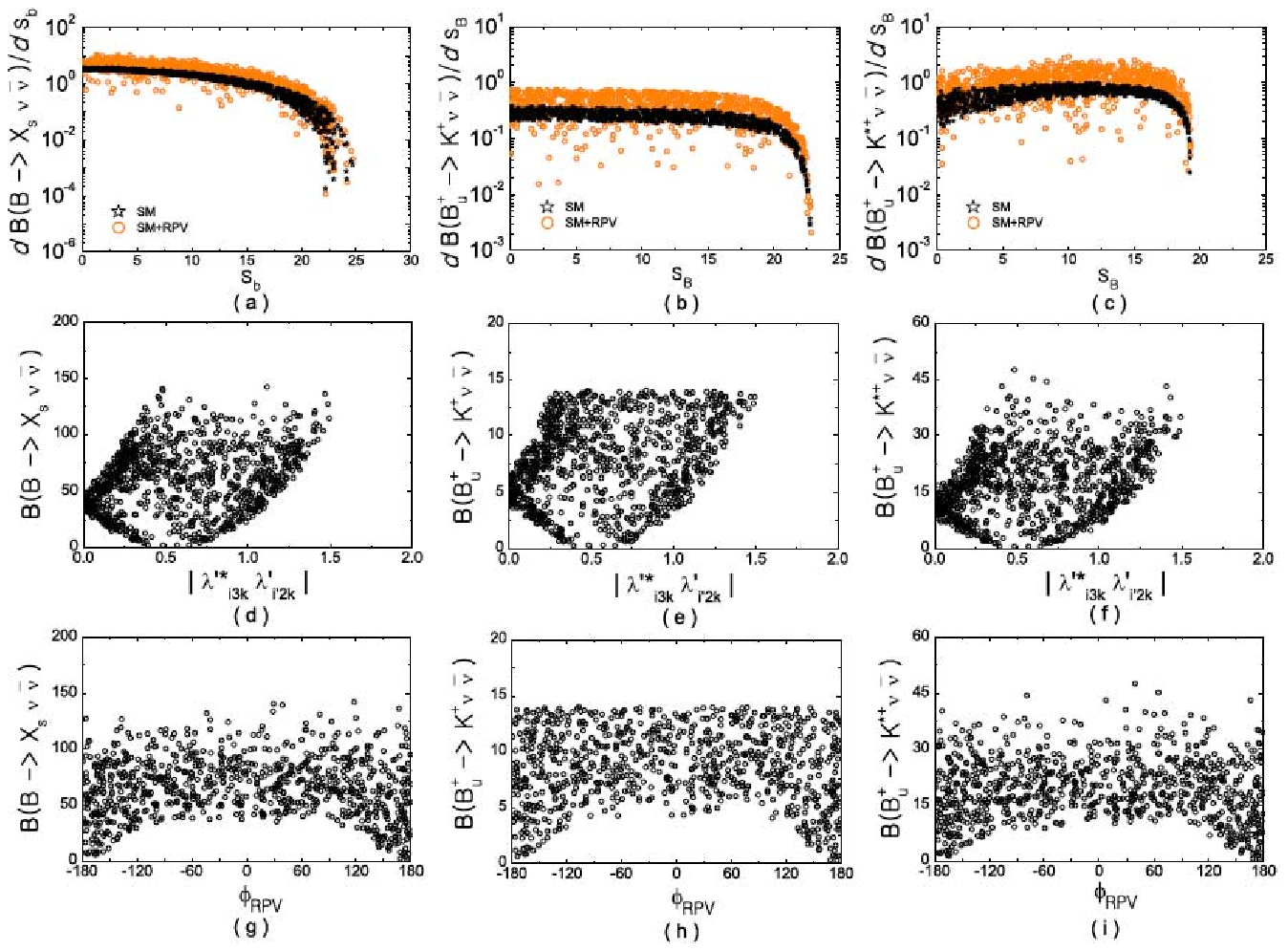}
\end{center}\vspace{-1cm}
\caption{\small The effects of RPV couplings
$\lambda'^*_{i3k}\lambda'_{i'2k}$ from left-handed squark exchanges
in $B\to X_{s}\nu\bar{\nu},K^+(K^{*+}) \nu\bar{\nu}$ decays.
$\mathcal{B}$ and $|\lambda'^*_{i3k}\lambda'_{i'2k}|$ are in units
of $10^{-6}$ and $10^{-2}$, respectively.} \label{Fig:sx}
\begin{center}
\includegraphics[scale=1]{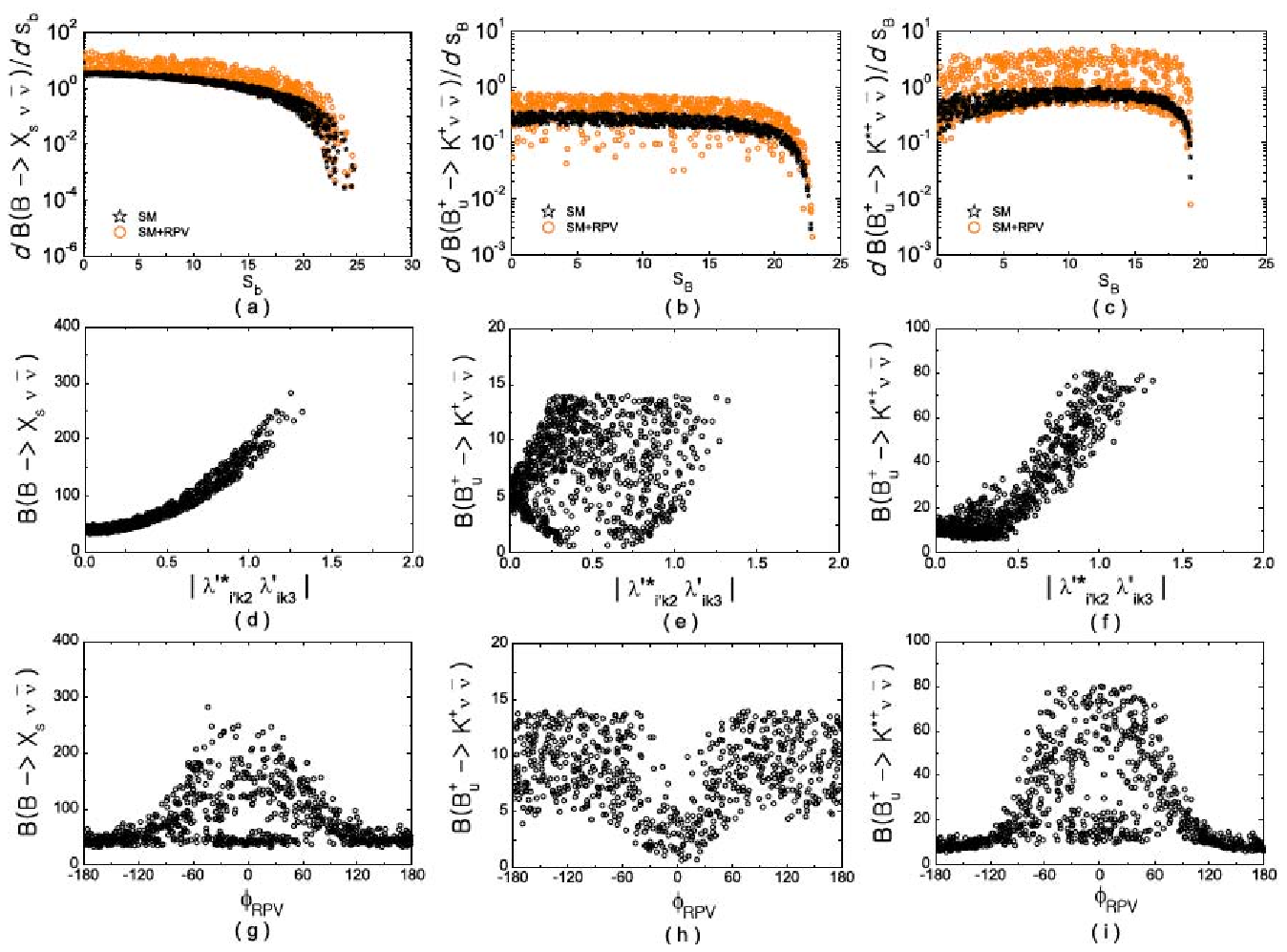}
\end{center}\vspace{-1cm}
\caption{\small The effects of RPV couplings
$\lambda'^*_{i'k2}\lambda'_{ik3}$ due to right-handed squark
exchanges in  $B\to X_{s}\nu\bar{\nu},K^+(K^{*+}) \nu\bar{\nu}$
decays. $\mathcal{B}$
 and
$|\lambda'^*_{i'k2}\lambda'_{ik3}|$ are in units of $10^{-6}$ and
$10^{-2}$, respectively.} \label{Fig:sy}
\end{figure}

\begin{figure}[t]
\begin{center}
\includegraphics[scale=1]{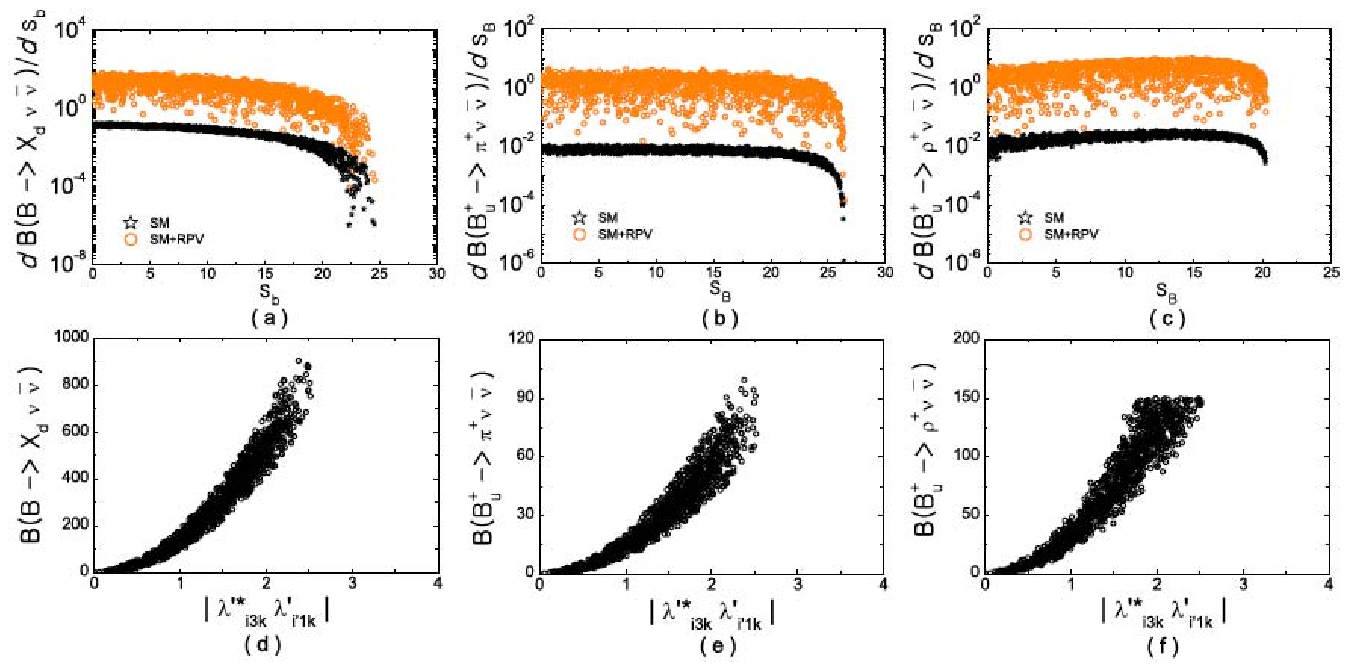}
\end{center}\vspace{-1cm}
\caption{\small The effects of RPV couplings
$\lambda'^*_{i3k}\lambda'_{i'1k}$ from left-handed squark exchanges
in $B\to X_{d}\nu\bar{\nu},\pi^+(\rho^+) \nu\bar{\nu}$ decays.
$\mathcal{B}$ and $|\lambda'^*_{i3k}\lambda'_{i'1k}|$ are in units
of $10^{-6}$ and $10^{-2}$, respectively.} \label{Fig:dx}
\end{figure}

 The effects of the RPV couplings
$\lambda'^*_{i3k}\lambda'_{i'2k}$ and
$\lambda'^*_{i'k2}\lambda'_{ik3}$ on $B\to X_{s}\nu\bar{\nu}$,
$K^{+}(K^{*+})\nu\bar{\nu}$ decays are shown in Fig. \ref{Fig:sx}
and Fig. \ref{Fig:sy}, respectively.
Now we turn to discuss plots of Fig. \ref{Fig:sx} in detail. Fig.
\ref{Fig:sx} displays the effects of RPV couplings
$\lambda'^*_{i3k}\lambda'_{i'2k}$ from left-handed squark exchanges
in $B\to X_{s}\nu\bar{\nu},K^+(K^{*+}) \nu\bar{\nu}$ decays. As
shown in Fig. \ref{Fig:sx}(a-c), $d\mathcal{B}(B\to
X_s\nu\bar{\nu})/ds_b$ and $d\mathcal{B}(B^+_u\to K^+ (K^{*+})
\nu\bar{\nu})/ds_B$  are obviously affected by
$\lambda'^*_{i3k}\lambda'_{i'2k}$ coupling, but the
$\lambda'^*_{i3k}\lambda'_{i'2k}$
 contributions to them
 cannot be distinguished from the SM expectations.
The scatter plots Fig. \ref{Fig:sx}(d-f) and Fig. \ref{Fig:sx}(g-i)
show $\mathcal{B}(B\to X_s\nu\bar{\nu}, K^+ (K^{*+}) \nu\bar{\nu})$
correlated with $|\lambda'^*_{i3k}\lambda'_{i'2k}|$ and its phase
$\phi_{_{\rm RPV}}$, respectively. From Fig. \ref{Fig:sx}(d-f),
 we see that $\mathcal{B}(B \to X_s\nu\bar{\nu}, K^+ (K^{*+})
\nu\bar{\nu})$ have some sensitivity  to
$|\lambda'^*_{i3k}\lambda'_{i'2k}|$, and they may have minima at
$|\lambda'^*_{i3k}\lambda'_{i'2k}|\approx 5\times10^{-3}$. Fig.
\ref{Fig:sx}(g-i) show that $\mathcal{B}(B\to X_s\nu\bar{\nu},K^+
(K^{*+}) \nu\bar{\nu})$ have high sensitivity to $\phi_{_{\rm RPV}}$
within $\mathcal{B}(B \to X_s \nu\bar{\nu})<40\times10^{-6}$ and
$\mathcal{B}(B^+_u\to K^+ (K^{*+})
\nu\bar{\nu})<5(10)\times10^{-6}$.
Fig. \ref{Fig:sy} shows RPV coupling
$\lambda'^*_{i'k2}\lambda'_{ik3}$ effects due to right-handed squark
exchanges in  $B\to X_{s}\nu\bar{\nu},K^+(K^{*+}) \nu\bar{\nu}$
decays.   {}From Fig. \ref{Fig:sy}(a-c), we can see the effects of
$\lambda'^*_{i'k2}\lambda'_{ik3}$ on $d\mathcal{B}(B\to X_{s}
\nu\bar{\nu})/ds_b$ and $d\mathcal{B}(B^+_u\to K^+ (K^{*+})
\nu\bar{\nu})/ds_B$  are very similar to those of
$\lambda'^*_{i3k}\lambda'_{i'2k}$ shown in Fig. \ref{Fig:sx}(a-c).
As shown in Fig. \ref{Fig:sy}(d,f,g,i), $\mathcal{B}(B\to
X_s\nu\bar{\nu}, K^{*+} \nu\bar{\nu})$ are also very sensitive to
$\lambda'^*_{i'k2}\lambda'_{ik3}$ coupling, and $\mathcal{B}(B\to
X_s\nu\bar{\nu}, K^{*+} \nu\bar{\nu})$ are obviously increasing with
$|\lambda'^*_{i'k2}\lambda'_{ik3}|$ but decreasing with
$|\phi_{_{\rm RPV}}|$. Fig. \ref{Fig:sy}(e,h) show that
$\mathcal{B}(B^+_u\to K^+ \nu\bar{\nu})$ is sensitive to
$\lambda'^*_{i3k}\lambda'_{i'2k}$ coupling, and it has minimum at
$|\lambda'^*_{i3k}\lambda'_{i'2k}|\approx 5\times10^{-3}$ and
$\phi_{_{\rm RPV}}\approx0^\circ$.

Since the branching ratios of the semi-leptonic $b\to d
\nu\bar{\nu}$ decays are not sensitive to the relevant RPV weak
phases, we will only show the correlations between the branching
ratios and the moduli.
Fig. \ref{Fig:dx} illustrates the contributions of RPV coupling
$\lambda'^*_{i3k}\lambda'_{i'1k}$ to
 $B\to X_d\nu\bar{\nu}, \pi^+
(\rho^{+}) \nu\bar{\nu}$ decays. As shown in the two-dimensional
scatter plots Fig. \ref{Fig:dx}(a-c),
$\lambda'^*_{i3k}\lambda'_{i'1k}$ coupling may change the order of
$d\mathcal{B}(B \to X_d \nu\bar{\nu})/ds_b$ and
$d\mathcal{B}(B^+_u\to \pi^+ (\rho^{+}) \nu\bar{\nu})/ds_B$, and the
$\lambda'^*_{i3k}\lambda'_{i'1k}$
 contributions to them
 are possibly distinguishable from the SM expectations at all $s_b(s_B)$ regions.
Fig. \ref{Fig:dx}(d-f) show $\mathcal{B}(B\to X_d\nu\bar{\nu},\pi^+
(\rho^{+}) \nu\bar{\nu})$ correlated with
$|\lambda'^*_{i3k}\lambda'_{i'1k}|$, and we see that
$\mathcal{B}(B\to X_d\nu\bar{\nu}, \pi^+ (\rho^{+}) \nu\bar{\nu})$
are greatly increasing with $|\lambda'^*_{i3k}\lambda'_{i'1k}|$.
The effects of RPV couplings $\lambda'^*_{i'k1}\lambda'_{ik3}$ due
to right-handed squark exchanges in  $B\to
X_{d}\nu\bar{\nu},\pi^+(\rho^+) \nu\bar{\nu}$ decays are very
similar to those of $\lambda'^*_{i3k}\lambda'_{i'1k}$ in these
decays shown in Fig. \ref{Fig:dx}, and we will not show the
correlations between observables and RPV coupling
$\lambda'^*_{i'k1}\lambda'_{ik3}$ again.

As we mentioned before, some of our combined bounds
are weaker one or two order(s) of magnitude than the existing
bounds. Now we are ready to discuss the RPV coupling effects after
also considering relevant previous stronger bounds. From above
analysis, we know that the left-handed squark exchange RPV couplings
have not evident effects on the branching ratios of $B\to
X_s\nu\bar{\nu},K^{+} \nu\bar{\nu}, K^{*+} \nu\bar{\nu}$ decays if
$|\lambda'^*_{i3k}\lambda'_{i'2k}|<2.7\times10^{-3}$.
$\mathcal{B}(B\to X_{s}\nu\bar{\nu},K^{*+} \nu\bar{\nu})$ will not
be obviously affected by RPV couplings due to right-handed squark
exchanges if $|\lambda'^*_{i'k2}\lambda'_{ik3}|<1.7\times10^{-3}$,
and $\mathcal{B}(B\to K^+ \nu\bar{\nu})$ will not be obviously
affected if $|\lambda'^*_{i'k2}\lambda'_{ik3}|<1.1\times10^{-3}$. As
for $B\to X_{d}\nu\bar{\nu},\pi^+(\rho^+) \nu\bar{\nu}$ decays, the
RPV coupling contributions   still can distinguish from the SM ones
if $|\lambda'^*_{i3k}\lambda'_{i'1k}|$ and
$|\lambda'^*_{i'k1}\lambda'_{ik3}|$ are larger than
$6.1\times10^{-4}$.  Then we can give a conclusion safely, the RPV
couplings  $\lambda'^*_{ik2}\lambda'_{ik3} ~(i=1,2)$, which moduli
are less than $1.2\times10^{-3}$ from $b\to s \ell^+\ell^-$
\cite{Xu:2006vk},  give small contributions to $B\to
X_s\nu\bar{\nu},K^{+} \nu\bar{\nu}, K^{*+} \nu\bar{\nu}$ decays, and
all other relevant couplings still can give remarkable contributions
to the  semi-leptonic $b\to (s,d)\nu\bar{\nu}$ decays after
considering the existing bounds.

\section{Summary}

In this letter we have performed a brief study of the RPV coupling
effects in supersymmetry from the exclusive and inclusive
semi-leptonic $B$ decays with a $\nu\bar{\nu}$ pair, which include
$B^+_u\to K^{(*)+}\nu\bar{\nu}$, $B^0_d\to K^{(*)0}\nu\bar{\nu}$,
$B^0_s\to \phi\nu\bar{\nu}$, $B\to X_s\nu\bar{\nu}$, $B^0_d\to
\pi^0(\rho^0)\nu\bar{\nu}$, $B^+_u\to \pi^+(\rho^+)\nu\bar{\nu}$,
$B^0_s\to K^{(*)0}\nu\bar{\nu}$ and $B\to X_d\nu\bar{\nu}$ thirteen
decay modes. Considering the theoretical uncertainties, we have
obtained conservatively constrained parameter spaces of RPV coupling
constants from the latest experimental upper limits.  We found, at
present, the strongest bounds on the relevant RPV couplings come
from the exclusive decays.  Furthermore, we also investigated the
sensitivities of the di-neutrino invariant mass spectra and
branching ratios to the survived R-parity violating coupling spaces.

 We have found that,  after satisfying all the current
experimental upper limits,  both left-handed and right-handed squark
exchange RPV couplings still have significant effects on these
di-neutrino invariant mass spectra and branching ratios. The RPV
contributions are not easily distinguishable from the SM predictions
in the di-neutrino invariant mass spectra of the semi-leptonic $b\to
s \nu\bar{\nu}$ decays, nevertheless, the di-neutrino invariant mass
spectra of the  semi-leptonic $b\to d \nu\bar{\nu}$ decays are very
useful to distinguish the RPV coupling effects at all kinematic
regions.
The branching ratios of the  semi-leptonic $b\to s \nu\bar{\nu}$
decays are sensitive to both  moduli and phases of relevant RPV
coupling products, and the branching ratios of the semi-leptonic
$b\to d \nu\bar{\nu}$ decays are only very sensitive to the moduli
of relevant RPV coupling products.

However, observing rare $B$ decays with a $\nu\bar{\nu}$ pair is
experimentally very challenging because of the two missing neutrinos
and (many) hadrons, and these decays can be searched for through the
large missing energy events in $B$ decays. With an advent of
Super-$B$ facilities \cite{Bona:2007qt}, the prospects of measuring
the branching ratios of the  semi-leptonic $b\to s \nu\bar{\nu}$
decays in next decade could be highly realistic, and it's also
possible to observe $B^+\to \pi^+\nu\bar{\nu}$ decay.  We expect
that future experiments will significantly strengthen the allowed
parameter spaces for RPV couplings. Our predictions of RPV effects
on related observables could be very useful for probing RPV
supersymmetric effects in future experiments.

\section*{Acknowledgments}

The work of C. S. Kim was supported
by the KRF Grant funded by the Korean Government (MOEHRD) No.
KRF-2005-070-C00030.  The work of Ru-Min Wang was supported by the
second stage of Brain Korea 21 Project.

\end{document}